\documentclass[sigconf]{acmart}

\usepackage{amsmath}
\usepackage{amsthm}
\usepackage{amsfonts}
\usepackage{graphicx}
\usepackage{hyperref}
\usepackage{setspace}
\usepackage{siunitx}
\usepackage{support-caption}
\usepackage{subcaption}
\usepackage{enumitem}
\usepackage{tikz}
\usetikzlibrary{positioning}
\usetikzlibrary{shapes}
\usetikzlibrary{shadows}
\usetikzlibrary{plotmarks}
\usetikzlibrary{decorations.pathreplacing}
\usepackage[export]{adjustbox}
\usepackage[ruled,vlined,linesnumbered]{algorithm2e}
\usepackage{pgfplots}
\pgfplotsset{compat=1.17}
\usepackage{ragged2e}

\AtBeginDocument{%
  \providecommand\BibTeX{{%
    \normalfont B\kern-0.5em{\scshape i\kern-0.25em b}\kern-0.8em\TeX}}}

\acmConference[LEAC]{}{Workshop on Learning in Control. CPS-IOT Week}{May 18, 2021} 
\renewcommand\footnotetextcopyrightpermission[1]{}

\begin{document}
\title{Automata Learning for Automated Test Generation of~Real~Time~Localization~Systems}

\author{Swantje Plambeck}
\email{swantje.plambeck@tuhh.de}
\affiliation{%
  \institution{Institute of Embedded Systems}
  \institution{Hamburg University of Technology}
  \streetaddress{Am Schwarzenberg-Campus 3}
  \city{Hamburg}
  \country{Germany}
  \postcode{21073}
}

\author{Jakob Schyga}
\email{jakob.schyga@tuhh.de}
\affiliation{%
  \institution{Institute for Technical Logistics}
  \institution{Hamburg University of Technology}
  \streetaddress{Theodor-Yorck-Straße 8}
  \city{Hamburg}
  \country{Germany}
  \postcode{21079}
}

\author{Johannes Hinckeldeyn}
\email{johannes.hinckeldeyn@tuhh.de}
\affiliation{%
  \institution{Institute for Technical Logistics}
  \institution{Hamburg University of Technology}
  \streetaddress{Theodor-Yorck-Straße 8}
  \city{Hamburg}
  \country{Germany}
  \postcode{21079}
}

\author{Jochen Kreutzfeldt}
\email{jochen.kreutzfeldt@tuhh.de}
\affiliation{%
  \institution{Institute for Technical Logistics}
  \institution{Hamburg University of Technology}
  \streetaddress{Theodor-Yorck-Straße 8}
  \city{Hamburg}
  \country{Germany}
  \postcode{21079}
}

\author{Görschwin Fey}
\email{goerschwin.fey@tuhh.de}
\affiliation{%
    \institution{Institute of Embedded Systems}
    \institution{Hamburg University of Technology}
    \streetaddress{Am Schwarzenberg-Campus 3}
    \city{Hamburg}
    \country{Germany}
    \postcode{21073}
}

\begin{abstract}
    Cyber Physical Systems (CPSs) are often black box systems for which no exact model exists. Automata learning allows to build abstract models of CPSs and is used in several scenarios, i.e. simulation, monitoring, and test case generation. Real time localization systems (RTLSs) are an example of particularly complex and often safety critical CPSs. We present a procedure for automatic test case generation with automata learning and apply this approach in a case study to a localization system.
\end{abstract}




\maketitle


\section{Introduction}
\label{sec:intro}

Many control systems are cyber physical systems (CPSs) that depend on complex physical effects. Most of the systems, in addition, are black box systems for which only the interaction with their surroundings, i.e., input and output characteristics are available while no knowledge of the inner working is given. Still, concise models of CPSs are needed for many applications, e.g., monitoring, maintenance, and test case generation. Some physical properties can be modeled by equations, in particular differential equations \cite{Mathmodel1,Mathmodel2}, but in most cases, many effects have to be considered at a time and exact models of a CPS including all its environmental influences do not exist.
Learning algorithms can help in modelling the complex dependencies in a CPS. 
Machine learning algorithms showed already good performance in modeling CPSs and other complex systems \cite{CPS-MachineLearn,CPS-MachineLearn2}.
The use of neural networks is widespread, but these models leak in their interpretability. In this paper, we consider automata as alternative models of a CPS \cite{murphy1995passively,Angluin87,active_learning}. The underlying assumption is that the functionality of the CPS can be abstracted to a finite state machine. 
An automaton model represents an input-output behavior of a system and, thus, is interpretable. An interpretable model generates knowledge about a system's functionality and allows the generation of complete, application-specific sets of test cases. We use the automaton model to perform an automated test case generation \cite{ABE+:2019}.

We apply our approach to real time localizations systems (RTLSs) for indoor localization. Localization systems are often proprietary and their inner working remains secret. 
RTLSs are measuring systems which have a strong interaction with physical effects of their environments like light conditions, temperature, surrounding structures, vibration, infrastructure, and more. These are typical characteristics of CPSs. Additionally, RTLSs are often applied in safety critical scenarios.
Our goal is to determine the relation between environmental influences and the localization quality. There are several environmental factors that might influence the localization quality but there does not yet exist a model for these influences. Our aim is to build interpretable models representing environmental influences while the approach is agnostic to the inner working. This includes an extensive testing of localization systems which can be supported by automated test case generation.

We present a work-flow for automated test case generation for black box systems and discuss critical aspects of this procedure within the case study of RTLSs. Automata learning is used to build an abstract model of the given system. Then, we apply algorithms for the test case generation. 
The procedure of model building and automated test case generation builds on established algorithms. The contribution of this work is the joined usage of established algorithms for a complete workflow in addition to discussions on critical aspects in this procedure.
Particularly important aspects are the feature selection for learning of abstract models and the algorithms for test case generation. 
Additionally, we present a concept for the learning and testing procedure and present a proof of concept with a localization system.

The paper is structured as follows: First of all, Section~\ref{sec:relatedwork} discusses abstractions of systems and presents previous applications of automata learning to CPSs or industrial systems. Next, we explain the theoretical background for automata learning and discuss the feature selection, the learning procedure, and methods for test case generation in Section~\ref{sec:model} and \ref{sec:test}. Afterwards, we describe in detail the experimental setup for a localization system in Section~\ref{sec:experiment}. In Section~\ref{sec:example}, we present some experimental results and discuss a simple model of a localization system. Then, Section~\ref{sec:future} gives an overview on the next steps for our approach. In the end, Section~\ref{sec:conclusion} concludes the paper.
\section{Related Work}
\label{sec:relatedwork}

Previous works extensively discussed the abstraction of complex, hybrid or cyber physical systems \cite{ABE+:2019,Piovesan2006,Shoham2008,Cassel2016}. In \cite{ABE+:2019} the need of a mapping between concrete data and abstract symbols, i.e., an alphabet for learning of an automaton is discussed. We readopt this issue and discuss a feature selection and data discretization. Still, it is not clear whether an abstract system has a valid automaton representation. This issue is investigated in \cite{Piovesan2006} which states theoretical limitations for the abstraction of hybrid systems to specific automata representations.

In \cite{online_passive}, automata learning is used for model building in an industrial scenario. The paper addresses different algorithms for automata learning and proposes an improved version of passive automata learning for complex systems. In contrast to test case generation, the model is used for monitoring of industrial production systems.

There exist approaches which apply automata learning for test case generation, e.g., for software \cite{Chow1978,Cicala2016} or hybrid systems \cite{ABE+:2019}. In \cite{ABE+:2019} automata learning and neural networks are combined to generate a model and test cases with the example of an autonomous driving scenario. The general procedure in this paper is similar to our presented approach because the work-flow bases on established techniques for model building and test case generation. Nevertheless, \cite{ABE+:2019} does not use the automaton representation to model the CPS but applies automata learning and test case generation only to generate a data set for learning of models with neural networks. This stays in contrast to our approach, searching for \textit{interpretable} models of a CPS. 
\section{Building Abstract Models for Test Generation}
\label{sec:model_test}
In this section, we describe the work-flow for automata learning and test case generation. First, we introduce to theoretical principles for the abstraction of a concrete black box system to an abstract system. The procedure, then, starts with data acquisition from the RTLS which results in a set of sequences of positioning data. Automata learning is not directly possible on this data, but a preparation step is required. The preparation step includes a selection of features, that allow an interpretable, abstract model, and a discretization of the data. With the prepared data, an abstract model of the system is learned. We apply algorithms for test case generation on the automaton model. Then, the abstract test cases have to be mapped back to test cases, i.e., test trajectories for the localization system.

\subsection{Learning an Abstract Model}
\label{sec:model}

\begin{figure}
    \includegraphics{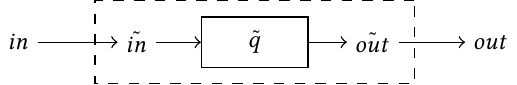}
    \caption{Abstract Model of the Localization System}
    \Description{A block diagram of the abstraction is shwon. The system, with concrete in- and outputs builds the inner functionality of the abstract system with abstract n- and outputs.}
    \label{fig:blackbox}
\end{figure}

The RTLS is a black box system for which an abstract model of the environmental dependencies is sought.
We assume an abstract model according to Fig.~\ref{fig:blackbox}. The solid rectangle represents the black box system, which receives inputs, e.g., trajectories and returns outputs, e.g., expected positions from the localization system. Additionally, the black box system has an unknown inner state $\tilde{q}$ as well as an unknown function $\tilde{out} = \tilde{\beta}(\tilde{in},\tilde{q})$ which together represent the real functionality of the system.
The dashed rectangle represents the abstract system, which receives abstract inputs, e.g., distances from obstacles, and returns abstract outputs, e.g., positioning errors of the localization system.
We assume that the abstract system can be modeled as a Mealy machine.

A Mealy machine $M$ is a tuple $(Q,I,O, \alpha, \beta, q_0)$, where
$Q$ is the set of states,
$I$ is the input alphabet, which here consists of all possible abstract inputs $in$,
$O$ is the output alphabet, which here consists of all possible abstract outputs $out$.
The functions $\alpha : I \times Q \rightarrow O$ and $\beta : I \times Q \rightarrow Q$ define the output and transition function, respectively. The initial state is $q_0 \in Q$ \cite{Automatatheory}.
The state $q \in Q$ and the functions $\alpha$ and $\beta$ approximate the inner behavior of the black box system, determined by $\tilde{\beta}(\tilde{in},\tilde{q})$ and $\tilde{q}$. The inner state $\tilde{q}$ in general follows a continuous behavior for CPSs and, thus, the mapping to a Mealy machine is only valid if a discrete abstraction of the inner working of the black box system is possible. But especially for infinite state spaces, automata learning is a suitable theoretical framework \cite{Angluin87,Piovesan2006}.





Our learning procedure starts from the black box system. We execute the mapping between the concrete signals $\tilde{in}$ and $\tilde{out}$, which are assumably continuous, to abstract signals $in$ and $out$, starting with a data acquisition of concrete samples from the continuous signals $\tilde{in}$ and $\tilde{out}$.
The concrete data samples from the RTLS are transformed to abstract samples in three steps:
\begin{enumerate}[topsep=0pt]
    \item selection of meaningful features,
    \item discretization of values from the selected features,
    \item reducing the data samples.
\end{enumerate}
Step (1) defines the information that is used for learning. This step directly influences the dependencies that are displayed in the learned model. We have to select at least one feature for the input signal $in$ and the output signal $out$. The learned model then represents the influence of the input feature on the output feature.
As automata learning is not possible on continuous numbers, the data from the selected features is discretized in Step (2).
This step is important because a low resolution or unsuitable borders for the discretization can lead to a loss in information.
Step (3) is optional. Reducing the amount of data samples is useful to emphasize characteristics for learning and/or to reduce noise in the collected data.
There are several options for the reduction of the data, for example sampling the original data, averaging over consecutive data, or merging successive, identical data.

After data acquisition and preparation, we use passive automata learning to generate a Mealy machine of the abstract system. Passive automata learning algorithms learn from a set of samples $S$. A sample $\mathbf{s} = \langle o_1, \dots, o_m \rangle \in S$ is a sequence of observations $o_i$, where each observation, is an input-output combination:\\ \mbox{$o_i = \langle in_i, out_i \rangle, in_i \in I, out_i \in O$}. 
The number of observations can differ between the samples.
One possible algorithm for passive automata learning is the RPNI algorithm \cite{Parekh1997}.
For the localization system, samples are generated from several measurements along random trajectories. A detailed description on the data acquisition for RTLSs is given in Section~\ref{sec:experiment}.
\subsection{Automated Test Generation}
\label{sec:test}

For automated test case generation, we rely on established algorithms which are described in the following. \cite{mbTesting} The output of the automated test generation are sequences of input features.

\textit{State coverage (SC):} State coverage returns a set of input sequences $S_{SC}$ which ensure that all states of the automaton are visited at least once, when applied to the system.

\textit{Transition coverage (TC):} Transition coverage returns a set of input sequences $S_{TC}$ which ensure that all transitions of the automaton are passed at least once, when applied to the system.


State coverage and transition coverage generate test cases automatically without any further information except for the learned automaton. State coverage does not cover all transitions and in general results in fewer test cases than transition coverage. 
Transition coverage, instead, makes sure that all transitions are used. Nevertheless, there may exist several paths for visiting all transitions. 

Test case generation could be improved by using prior knowledge on the systems, i.e. about the meaning of input-output combinations (input-output combinations that are critical for the system could be preferred) or about the frequency with which transitions are used within the sample data.


The result of the test case generation is a set of sequences of input features. These input features represent abstract and/or discretized influences to the system. Thus, they do not directly represent a test case that is applicable to the system under test and a mapping between input sequences and applicable test cases is needed.
\section{Experimental Results}
\subsection{Experiment Description}
\label{sec:experiment}

There exist several different kinds of RTLSs.
The presented results are achieved with a Light Detection and Ranging (LiDAR)-system.
The measurement is based on the \textit{time of flight} method. The distances from the sensor to the contours of the environment are determined based on the time between emission and detection of a reflected laser impulse \cite{DBLP:journals/corr/CadenaCCLSN0L16}.
The measurement takes place in several scanning layers with a field of view of 180°.
Comparing the measurements to a prerecorded map
yields in an estimated position and orientation of the sensor. 
Examples for expectable influences on the quality of the localization are the spatial environment and its dynamics, the map quality, the lightning conditions, or the sensor's movement.
Since the given system is proprietary with highly complex inner workings and unknown error propagation, it is considered a black box.

The experiment was conducted at the testing facility of the TUHH Institute for Technical Logistics on a rectangular area of \SI{80}{\meter\squared}.
To gain the ground truth position of the entity under localization, an optical motion capture system is installed \cite{article_mocap}. This reference system later allows us to determine the localization error of the RTLS.
The localization sensor is attached to a trolley, which is manually pushed along a predefined trajectory, i.e. a path on the rectangular testbed.
The position information is received with a frequency of \SI{20}{\hertz}. Time synchronization between the systems is achieved by using the precision time protocol.

One run of the experiment results in a set of output data consisting of 5000 to 15000 pairs, depending on the trajectory length of the experiment run, of absolute positions and orientations measured by the localization and the reference system.
From this data many further quantities can be derived, such as:
\begin{itemize}[topsep=0pt]
    \item position and position error (along axis, euclidean error, etc.)
    \item movement direction and orientation error
    \item position and orientation covariances
    \item distances from objects of the spatial environment.
\end{itemize} 

The absolute euclidean position error, which is a main metric to describe the localization quality, is chosen as the output feature \cite{isonorm}.
The results suggest strong influences from the sensor's orientation on the euclidean error. 
This effect can be explained by the sensor's field of view facing a certain side of the facility. Therefore, the orientation is chosen as the input feature as a first learning approach.

To determine whether the learned model is a valid model, a validation step is applied after learning. We divide the set of data samples from the experiments into a learning set and a validation set prior to learning with a ratio of approximately 2:1. The model is then built based on the learning set. Afterwards, we validate whether the model is conform with the samples from the validation set. 
The share of samples from the validation set, that are valid on the model, is called the \textit{accuracy}. 
If the model represents the validation set at least with a required minimal accuracy, the learning process is done. Otherwise, further samples are needed for learning. 

\subsection{Experimental Validation - Simple Examples}
\label{sec:example}



\begin{figure}[th]
    \includegraphics{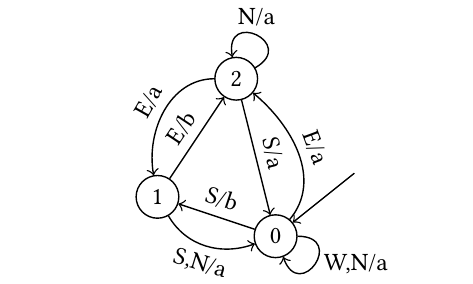}
    \caption{Simple Exemplary Automaton of a Localization System}
    \Description{An automaton with 3 states is shown. A loop between the three states with the anotations S/b, E/b, S/a exists.}
    \label{fig:simple_exmp}
\end{figure}

As stated in the previous section, we chose the localization error as an output feature and the orientation as an input feature for automata learning.
The localization error is discretized to two classes which we call $a$ and $b$, where $a$ represents non-critical errors and $b$ represents critical errors. The orientation on the other hand is discretized in 4 sectors (north (N), east (E), south (S), west (W)). The sectors face a corresponding side of the rectangular testbed. Additionally, successive, identical data are merged to one observation, i.e., a sample 
\mbox{$\mathbf{\tilde{s}_k} = \langle \langle N,a\rangle, \langle N,a\rangle, \langle N,a\rangle, \langle E,b\rangle,\langle E,b\rangle, \langle N,a\rangle \rangle$}
is reduced to 
\mbox{$\mathbf{s_k} = \langle \langle N,a\rangle, \langle E,b\rangle, \langle N,a\rangle \rangle$}.
A raw data set of approx. 10000 data points is reduced to a sample with less then 100 observations, which is sufficient to learn a small automaton model, by this manner.
From this single sample, we get a small and simple automaton model which is shown in Fig.~\ref{fig:simple_exmp}. 

The automaton has three states $Q = \{0,1,2\}$, four possible input signals $I = \{N,E,S,W\}$ and two possible output signals $O = \{a,b\}$. The initial state is the state $q_0 = 0$.
If we perform a state coverage algorithm on the system, we get the input sequences \mbox{$S_{SC}=\{\langle N \rangle,\langle S \rangle,\langle S,E \rangle\}$} as a possible result. 
Transition coverage could result in the following set of sequences: \\
\mbox{$S_{TC} = \{\langle N \rangle,\langle E,N\rangle, \langle E,E\rangle, \langle E,S\rangle, \langle S,S\rangle,\langle S,E\rangle\}$}.
Each member of the sets $S_{SC}$ and $S_{TC}$ gives a sequence of inputs that result in a test case of the abstract system. For example, the sequence $\langle S,E \rangle$ corresponds to a test case, where the system is first oriented in the southern direction and then in the eastern direction. 



The generated test cases represent sequences of orientations in the horizontal plane. As a simple rotation at a single position is not a useful test scenario for the localization system, we map each orientation to a movement of a unit distance in the direction of this orientation. This maps, for example, $\langle S,E\rangle$ to the test case '\textit{Move one Meter South, Move one Meter East}'.

\begin{table}
    \small
\renewcommand{\arraystretch}{0.9}
\begin{tabular}{r|r|r|r|r}
    $|I|$ & $|O|$ & accuracy & $|S_{SC}|$ & $|S_{TC}|$ \\
    \hline
    4 & 2 & 1.0 & 6 & 25\\
    6 & 2 & 0.67 & 8 & 43\\
    8 & 2 & 0.33 & 8 & 54\\
\end{tabular}
    \caption{Exp. Results for 5 Learning and 3 Validation Samples}
    \label{tab:exp_res}
\end{table}

Table~\ref{tab:exp_res} shows results on further experiments with changed discretization, i.e., sectors of the orientation, with abstract samples of lengths between 10 and 100. With increasing size of the input set, the model size increases and also the number of test cases resulting from the automated test generation increases. For larger models, more samples are required to build exact models. Thus, we observe smaller accuracy on the validation set for these cases.


\section{Future Work}
\label{sec:future}


So far, we assumed that the abstraction to a Mealy machine is valid for our abstraction of the localization system without a prove. For the presented simple models which base on a very limited input alphabet, it is likely that a discrete abstraction of the inner working is valid. We want to determine characteristics or limitations to systems and abstractions that assure a valid abstraction also for more complex scenarios.

The process of model building allows the identification of relevant influences. We plan to use an automatic process to decide for relevant features from a list of possible influences. Part of this approach is also the automatic selection of meaningful discretizations.

Regarding test case generation the usage of more advanced techniques is possible and also the usage of prior knowledge in addition to the model is possible.

For the example of the localization system, it is planned to include more features to the learning process. Especially, influences like light conditions or interfering signals will be considered in addition to the purely position- or orientation-based influences. The goal is to gain a model that is independent from a concrete environment but holds globally for a localization system. 
Furthermore, for automata learning multiple features should be considered simultaneously as inputs to the system.

Apart from these ideas, we want to consider new approaches for model building. Automata learning aims for the construction of exact models, but for real-world applications it might be useful to allow approximate models to avoid overfitting. We propose to use simple algorithms like decision trees for the construction of abstract models and compare these approaches to automata learning.
\section{Conclusion}
\label{sec:conclusion}

In this paper, we presented a complete work-flow for the generation of abstract models from a CPS together with automated test generation using automata learning. The procedure consists of the steps: data acquisition, data preparation (including feature selection and discretization), automata learning for model building, and test case generation together with a mapping from abstract test cases to applicable test cases.
We claim that the feature selection and the test case generation are crucial points of this process and identify potentials to improve these steps regarding the generation of models for CPSs.
The presented work-flow is applied for the testing of localization systems as a proof of concept.

\begin{acks}
    This work is supported by the TUHH $I^3$ Projects funding. Special thanks go to the 3D\_Log project team. 
\end{acks}

\begin{spacing}{0.85}
    \footnotesize
    \bibliographystyle{IEEEtranN}
    \bibliography{bib/swantje,bib/jakob}
\end{spacing}

\end{document}